\documentstyle[12pt]{article}
\begin{document}
\newcommand{\dir}{/ \! \! \! \! D}
\newcommand{\beq}{\begin{equation} }
\newcommand{\eeq}{\end{equation} }
\newcommand{\bea}{\begin{eqnarray} }
\newcommand{\eea}{\end{eqnarray} }
\newcommand{\eno}[1]{(\ref{#1})}
\sloppy
\begin{titlepage}
\samepage{
\setcounter{page}{0}
\rightline{DFTT-55/93}
\rightline{hep-ph/9310238}
\vspace{.1in}
\rightline{October 1993}
\vspace{.1in}
\begin{center}
{\Large\bf The light-flavor structure of the
nucleon\\}
\vspace{.3in}
{\large Stefano Forte\footnote{
Address after December 1, 1993: CERN, CH-1211 Gen\`eve 23, Switzerland}\\}
\vspace{.15in}
{\it I.N.F.N., Sezione di Torino,\\
via P.~Giuria 1, I-10125 Torino, Italy\\}
\end{center}
\vspace{1in}
\begin{abstract}

Recent data on the Gottfried sum and less recent ones on the pion-nucleon sigma
term seem to disagree with naive parton-based
expectations on the light (i.e., up, down and strange) quark content
of the nucleon. We show that these discrepancies are resolved if
nonperturbative contributions are included in the analysis of the data. These
appear both in the computation of matrix elements of operators, and in their
QCD scale dependence, and depend strongly on the quantum numbers of the given
state.

\end{abstract}
\vspace{.65in}
\begin{center}
{\it Presented at the XXXIII Cracow School of Theoretical Physics\\
Zakopane, Poland, June 1993\\}
\vspace{.1in}
{To be published in the proceedings}
\end{center}
\vfill}
\end{titlepage}

\section{Introduction}
In recent times a set of high-precision measurements of the nucleon structure
functions, both in the polarized and unpolarized case, has lead to results in
striking disagreement with naive parton model expectations. The
first of these to come \cite{EMC} has spawned a vast theoretical
literature\cite{spinrev}, the
so-called nucleon spin crisis;
it is now clear \cite{forspin} that the experimental
result can be fully explained and understood only in a nonperturbative
framework, and, more specifically, that it may be a first hint of a
nonperturbative effect which survives the high-energy limit and is thus visible
in the deep-inelastic regime.

More recent results on the
Gottfried sum rule \cite{NMCa}\cite{NMCb} have led to question
even the standard \cite{SSV} assumptions on the up and down (unpolarized)
flavor content of the nucleon. Combined with the long-standing sigma term
puzzle\cite{sigpuz}, which hints to an unusual strange quark content,
these results seem to suggest that the light-flavor structure of the
nucleon is rather different from what the naive intuition based on a
combination
of the
non-relativistic quark model with the naive parton model would
suggest.

It seems thus that naive expectations have to be reconsidered.
These are typically expressed as Ans\"atze on the form of the quark
distributions $q_i^k(x)$, which in the parton model provide the probability of
a quark of flavor $i$ to carry a fraction $x$ of the momentum of the baryon
$k$, and are related to
the nucleon structure functions.
The quark distributions are
split into a valence and a sea part:
\beq
q^b (x) = q^b_v (x) + q^b_s (x) \ ; \ \ \ \bar{q}^b (x) = q^b_s (x).
\label{naive}
\eeq
In QCD this decomposition has only meaning as a definition: the quark
distributions are related to the expectation value of certain operators in such
a way that the structure function $F_2(x)$ is related to them by\cite{altrev}
\beq
F_2^k(x)=x\sum_i e^2_i \left(q_i^k(x)+\bar q_i^k(x)\right),
\label{qandf}
\eeq
where the sum runs over all the quark flavors and $e_i$ are electric or weak
charges according to whether the target is probed with electrons or neutrinos.
The quark and antiquark components for various flavors can be
extracted by taking suitable linear combinations of structure functions for
neutrino, antineutrino and electron scattering off different targets.
QCD then leads to predict the scale dependence of these quantities.

However, it is usually assumed that the quark distributions can be actually
viewed as expressing the quark content of the given hadron, and that this
picture provides information on the nucleon structure in that it
suggests the form of the distribution themselves. This assumption
is based on the expectation that the matrix
elements of operators which
provide the quark-parton distributions (and which can only be written
down in the light-cone formalism \cite{soper}) can be identified, at least
up to
some unknown renormalization, with the quark operators
of the naive quark
model. Such an assumption is totally unjustified
within  QCD; it is however quite successful phenomenologically,
and it is customarily used in analyzing the experimental data.

A simple example of application of this wisdom is the computation of the ratio
$R(x)=F_2^n(x)/F_2^p(x)$ as $x\to1$.
The naive argument goes as follows: as $x\to1$
only the valence quarks survive, because if a quark is carrying the whole of
the nucleon's momentum it must be a valence one. Furthermore, if
one single quark is carrying the quantum numbers of a nucleon, by
isospin, it ought to
be a $u$ quark for the proton and a $d$ quark for the neutron.
Thus $\lim_{x\to1}R(x)=e^2_d/e^2_u=1/4$. This prediction is
spectacularly borne out by the
data \cite{NMCa},\cite{NMCrat}, while it is clearly based partly on symmetry
(isospin) and partly on naive guesswork.

More in general, the quark distributions \eno{naive} are usually taken to
satisfy the following
assumptions:
\begin{enumerate}
\item
The sea and valence components separately satisfy isospin symmetry:
\beq
 u^p_v = d^n_v, \quad  u^n_v = d^p_v; \qquad
u^p_s = d^n_s, \quad  u^n_s = d^p_s.
\label{ispin}
\eeq
\item
The sea component in each baryon is SU(2) symmetric:
\beq
u^p_s = d^p_s;\quad  \ u^n_s = d^n_s \label{qspin}.
\eeq
\item
The content of Zweig suppressed flavors, such as strange quarks in the nucleon,
is small. This
implies relations like
\beq
\langle p | \bar{s} s | p \rangle \approx
\langle n | \bar{s} s | n \rangle \approx 0.
\label{sigpeq}
\eeq
\end{enumerate}

It turns out that the recent data on the Gottfried sum rule put
either (or both) assumptions 1 and 2 into question, and the long-standing sigma
term puzzle challenges assumption 3.
In the sequel, we will discuss both these points in turn. We shall show that
the naive picture can actually be recovered, provided
nonperturbative QCD effects are taken into account, both in the QCD evolution
and in the determination of the symmetry structure of  matrix elements of
operators.

\section{The puzzle of the Gottfried sum}

The question of the validity of assumptions 1 and 2
can be condensed into an analysis of the
Gottfried sum $S_G$, defined by
\beq
S_G \equiv \int dx \, \frac{F_2^p(x) -F_2^n (x)}{x}.
\label{gotsum}
\eeq
According to Eq.\eno{qandf},
the first moment of $F_2 $ corresponds
to the sum of the electric charges squared of all partons present in the
target,
\beq
\int dx \, \frac{F_2 (x)}{x} = \sum_{i} e_i^2 \int dx \,
(q_i (x) + \bar{q}_{i} (x)) \equiv \sum_{i} e_i^2 n_i,
\eeq
where we have defined
\beq
n_i=\int\!dx\,\left(q_i(x)+\bar q_i(x)\right).
\label{nandq}
\eeq
This is expected to be  a divergent quantity for a single nucleon, due to the
fact that at small $x$ the nucleon's momentum is subdivided among an infinite
number of quarks; however this divergence (i.e., the small-$x$ behavior of
all $q^i(x)$) is expected to be universal
(on the basis of
Regge theory)\cite{leapre},
essentially due to the flavor-independence of the strong
interaction, hence $S_G$ Eq.\eno{gotsum} is expected to be finite.

The symmetry assumptions (1),(2) amount to the further
conjecture that not only in
the $x\to0$ limit, but actually for all $x$ the sea is universal, hence
only valence quarks can contribute to nonsinglet combinations such as
Eq.\eno{gotsum}. Taking only the valence component into account one
arrives thus at the so-called Gottfried sum rule\cite{leapre}\cite{GSR}:
\bea
S_G &=& \frac{4}{9} n_{u,v}^{p} + \frac{1}{9} n_{d,v}^{p} -
\frac{4}{9} n_{u,v}^{n} - \frac{1}{9} n_{d,v}^{n} \\
&=& \left( \frac{4}{9} -\frac{1}{9} \right) (n_{u,v}^{p} -n_{d,v}^{p} )
=\frac{1}{3}.
\label{naigot}
\eea
However $S_G$ has now been measured\cite{NMCa} as
\beq
S_G = 0.24 \pm 0.016 \mbox{ at}\> 4\;\mbox{ GeV}^2.
\label{gotexpa}
\eeq
 Even though
 the extraction
of this value involves certain assumptions, e.g. an extrapolation of the
measured
structure function to the full range of $x$, it seems exceedingly unlikely that
a value compatible with the naive expectation Eq.\eno{naigot} could be
obtained. As a matter of fact,
an improved data analysis \cite{NMCb} has given the new value
\beq
S_G = 0.258 \pm 0.017 \mbox{ at}\> 4\;\mbox{ GeV}^2.
\label{gotexpb}
\eeq
but has also shown evidence for shadowing in deuterium.
If shadowing effects (which cannot be measured precisely as yet)
were included, the value of $S_G$ as currently determined
(Eq.\eno{gotexpa} or \eno{gotexpb}) should be brought down by an
amount which is theoretically estimated\cite{shad} to be
$\Delta S_G\sim 0.2-0.4$. The value Eq.\eno{gotexpa} seems thus a
conservative estimate of the value of $S_G$, which might turn out to be
even lower.

It is thus necessary to reconsider assumptions 1 and 2. If we make no
assumption,
the
Gottfried sum is generally given by
\beq
S_G = \frac{1}{2} \left[ \frac{5}{9} (n_u +n_d ) + \frac{1}{3}
(n_u -n_d ) \right]_{I=1}
\label{gotgen}
\eeq
where the subscript indicates that the isotriplet part must be taken,
i.e. the difference of the expression in square brackets between a
proton and a neutron. Assumption 1 (isospin symmetry) implies
$ [ n_u +n_d ]_{I=1}=0 $, whereas assumption 2 (sea flavor symmetry)
implies
that $ [ n_u -n_d ]_{I=1} $ be identified with its valence value,
i.e., $ [ n_u -n_d ]_{I=1}=1 $. The data
could be explained either by assuming that the neutron sea is bigger than the
proton one (thus violating isospin), or that the proton sea and separately the
neutron sea are not flavor neutral, but rather they have a flavor asymmetry
that anticorrelates to that of the valence component (thus violating sea flavor
symmetry).

In order to fix the relative amount of violation of these two symmetries
one needs an independent measurement of a different linear combinations of the
quantities entering Eq.\eno{gotgen}.
It can be shown\cite{gotfor}\cite{forsig}
that the required information is provided by
the isotriplet part $\sigma_G$ of
the so-called nucleon $\sigma $-term, defined as
\beq
\sigma_G\equiv\sigma_{I=1}=\langle N | m_u \bar u u + m_d \bar d d |
N\rangle_{I=1},
\label{sigg}
\eeq
where $m_i$ are  current quark masses. This is because $\sigma_G$ satisfies
\cite{gotfor}\cite{forsig}
\beq
\sigma_{G} = \frac{1}{2} \left[ (m_u +m_d ) (n_u + n_d ) +
(m_u -m_d ) (n_u - n_d ) \right]_{I=1} = M_p -M_n
\eeq
which is the analogue of the Gottfried sum $S_G $, with square charges
replaced by masses. Using this extra piece of information one gets
to the conclusion that
\bea
\frac{1}{2} [ n_u + n_d ]_{I=1} & \approx & \pm 0.03
\label{ibr}
\\
\frac{1}{2} [ n_u - n_d ]_{I=1} -1 & \approx & -0.35, \pm 0.1
\label{qbr}
\eea
i.e., sea flavor neutrality is broken considerably more strongly than
isospin.

Just as in the well-know case of the ``spin puzzle'' \cite{EMC}\cite{spinrev}
it turns out that the experimental result Eq.\eno{ibr}-\eno{qbr}
can be easily accommodated in effective models of the nucleon, such as the
Skyrme or the bag model\cite{gotfor}. Thus, again like in the spin case, one
may ask whether the data are not simply telling us that the ``naive parton
wisdom" is simply not applicable in this channel.

To make this more precise,
one should first however investigate what QCD does tell us.
Any nonsinglet combination of first moments of structure functions
$F_2(x)$, such as $S_G$ Eq.\eno{gotsum} acquires a two-loop
scale dependence via the
Altarelli-Parisi equations\cite{altrev}. Indeed, it is clear that at one loop
the quark number does not
evolve, because at one loop the only elementary process
(in the nonsinglet channel) is
gluon radiation by a quark line (Fig.~1) which clearly leaves the total
number of quarks (albeit not their momentum distribution) unchanged.
At two loops, however, the emitted gluon can in turn produce a quark-antiquark
pair through the diagram of Fig.~2, thereby increasing the total quark plus
antiquark number measured by the first moment of $F_2(x)$. At first sight it
would seem that this again should not contribute to the evolution of nonsinglet
quantities, because the probability for pair emission is independent of the
emitted flavor. However, this is not entirely correct because if the emitted
pair has the same flavor as the original quark the final state must be
antisymmetrized with respect to the two identical quarks, while this is of
course not the case if the flavor of the radiated pair is
different\cite{rosac}.
This is enough to generate a tiny difference between the splitting function
${\cal P}^D_{qq}$ for flavor-diagonal emission (i.e., the case when the
emitted pair
has the same flavor as the original quark) and the splitting function
${\cal P}^{ND}_{qq}$ for flavor-nondiagonal emission (i.e., the case when the
emitted pair
has a different flavor).

It follows that $S_G$ acquires a scale dependence\cite{rosac}; this is given by
the two-loop
Altarelli-Parisi equation\cite{altrev}
\beq
{d\over dt}(q\pm\bar q)^{NS}=
(q\pm\bar q)^{NS}\otimes \left(Q_{qq}\pm Q_{q\bar q}\right),
\label{tlap}
\eeq
where $\otimes$ indicated convolution of the splitting function
and the quark
distributions, $NS$ indicates
any nonsinglet combination of quark distributions,
and the splitting function responsible for the nonsinglet
evolution is given by
\beq
Q={\cal P}^D-{\cal P}^{ND}
\label{Qdef}
\eeq
in terms of the diagonal and nondiagonal splitting functions.
It follows that the evolution of $S_G$ Eq.\eno{gotsum}
is given by
\beq
{d\over dt}S_G(t)= \left(Q^1_{qq}+Q^1_{q\bar q}\right)S_G(t),
\label{tlapsg}
\eeq
in terms of the first moments  $Q^1_{qq}$,
$Q^1_{q\bar q}$ of the various splitting functions.
The physical
interpretation of Eq.\eno{tlap}\eno{tlapsg}
(which is derived e.g. in
Ref.\cite{altrev})
is straightforward: the nonsinglet evolution
is due to the difference in probability for flavor-diagonal and
flavor-nondiagonal pair production; if the latter is greater than the former
then the
nonsinglet quark number measured by $S_G$ decreases because the sea generated
from the process of Fig.~2 tends to compensate the starting flavor asymmetry.

The same results can be obtained \cite{rosac}
via operator-product expansion methods, which
are however much more cumbersome because only even moments of the
(charge-conjugation even) combination of structure functions which appears in
Eq.\eno{gotsum} have a direct interpretations in terms of matrix elements of
twist-two local operators, whereas the odd moments (such as the first moment
which
defines $S_G$) have to be constructed by analytic continuation.
Anyway, $S_G$ is thus found to evolve perturbatively according to
\beq
S_G (Q^2 ) = \left[ 1+\frac{\gamma^{(2)} }{b}
\left(\alpha_{s} (Q^2 ) -\alpha_{s} (\mu^{2} ) \right)\right] S_G (\mu^{2} ),
\label{tlsg}
\eeq
where explicit perturbative computation up to two loops leads to
$\gamma^{(2)} /b = 0.01 $
\cite{rosac}. It follows that perturbative QCD evolution, even though present,
is entirely negligible from a quantitative point of view because if $\alpha_s
< 1$ (as it must in order for perturbation theory to make sense) then the total
variation of $S_G$ is less than 1 \%.

It seems thus that, just like in the spin case, even if we assume that the
``parton wisdom'' should apply at some low scale, while the experimental result
is taken at large $Q^2$,
perturbative QCD evolution cannot explain the discrepancy between the
experimental value Eq.\eno{gotexpa} and the expectation Eq.\eno{naigot}.
More specifically, even if we are willing to give up the naive
expectation Eq.\eno{naigot} we must accept the rather unpalatable conclusion
that even at very
low scales the quark-parton content of a nucleon is very different
from its constituent quark content.

There is, however, a way out of  this impasse, which consists of allowing for
the possibility that the perturbative evolution Eq.\eno{tlsg} be replaced by
some nonperturbative evolution at low enough scales. Due to a
nonperturbative mechanism this evolution could then be much larger than
the perturbative one.
Symmetry considerations suggest what the origin of such a mechanism may be
\cite{ehq}.
If the interaction which governs the generation of the quark-antiquark
sea is U(N$_f$)-symmetric, then by Wigner-Eckhart's theorem the sea will be
flavor symmetric. In such case there is no QCD evolution at all, and indeed
Eq.\eno{tlap} shows that all  non-singlet scale dependence can be traced to a
difference in diagonal and non-diagonal splitting functions which signals the
violation of exact U(N$_f$) symmetry. The evolution is thus, crudely
speaking, a measure of the amount of U(N$_f$) symmetry breaking.
Now, it is well known that
in QCD  the axial
U$_A$ (N$_f$) flavor symmetry is dynamically broken down to a
SU$_A$ (N$_f$)  by the axial anomaly. This is  manifested
in the mass spectrum of pseudoscalar mesons, since
the singlet meson (the $\eta^{\prime }) $ has a much larger mass
than the octet mesons (pions and kaons)  which can be viewed as Goldstone
bosons. If such effect is incorporated in the Altarelli-Parisi
equations it
could thus potentially
lead to large evolution. Notice that this can only be done in
an effective way, since there is no known way of obtaining this pattern of
symmetry breaking through purely perturbative computations (rather,
nonperturbative effects such as instantons are thought to be at the origin of
it).

We can thus try to effectively include nonperturbative symmetry-breaking
effects by considering contributions to the
evolution equations where a meson couples directly to the quark-partons.
These can be viewed pictorially as being given by diagrams with the same
topology as those of Fig.2, but where the original quark and the emitted
antiquark exchange many gluons in such a way that they can be viewed as a
quark-antiquark bound states, thus leading to the diagram depicted in Fig.~3.
It is clear that, qualitatively, this should give a contribution to the
Altarelli-Parisi evolution which may potentially generate significant evolution
with the right sign as required to explain the observed effect
Eq.\eno{gotexpa}.
Indeed, due to the larger mass of the $\eta^{\prime } $, the
singlet meson emission should be
energetically suppressed in comparison
to octet meson emission. Because flavor-nondiagonal emission is entirely due to
the latter process, whereas flavor-diagonal emission proceeds in part through
the former, one expects that the diagonal process will be disfavored.
If Eq.\eno{tlap} may be generalized to meson emission, this
should result  in  a negative value of its r.h.s.,
thus leading to a partial screening of the valence flavor asymmetry.
Notice that the process given by the elementary diagram of Fig.3 is now
effectively a one-loop one, thus the corresponding evolution is potentially
strong.

It seems thus reasonable to investigate this option in a quantitative way
\cite{bafogo}.
The required generalization of the Altarelli-Parisi evolution proceeds in
two steps: first, we must generalize the set of evolution equations so
as to include the coupling of quarks to bound states, and then we must
compute the new splitting functions which enter the generalized equations.
Because we are only interested in nonsinglet evolution, we may concentrate on
the pseudoscalar meson
sector, since for all other mesons or baryons the flavor
symmetry will be approximately exact, thus, by the above  argument, no
contributions to the evolution should arise.

The main difference between the usual Altarelli-Parisi equations and the
generalized ones is then the possibility of flavor-nondiagonal transitions
already at the one loop level, such as that given by the diagram of Fig.~3 when
one starts with, say, a $u$ quark and ends with a $d$ quark and a  $\pi^+$.
It is then easy to see that the nonsinglet evolution equations generalize
(considering for simplicity flavor SU(2) only)
as
\bea
{d\over dt} q&=&  {\cal Q}_{qq} \otimes q
+{\cal Q}_{q\bar q}\otimes \bar q+ {\cal P}_{q\pi}\otimes\pi,
\label{qev}\\
{d\over dt} \bar q&=&  {\cal Q}_{qq} \otimes\bar q
+  {\cal Q}_{q\bar q}\otimes q- {\cal P}_{q\pi}\otimes \pi,
\label{qbarev}\\
{d\over dt} \pi&= & {\cal P}_{\pi q} \otimes (q-\bar{q})
+ {\cal P}_{\pi\pi} \otimes\pi
\label{piev},
\eea
where $q$, $\bar q$ and $\pi$ are quark, antiquark and pion nonsinglet
distributions , respectively, defined as
\bea
q(x)&=& q_u(x)-q_d(x)
\label{qdef}\\
 \bar{q}(x)&=&\bar{q}_u(x)-\bar{q}_d(x)
\label{qbardef} \\
\pi(x;t)&=&\pi^+(x;t)-\pi^-(x;t),
\label{pidef}
\eea
and, at one loop, the splitting functions ${\cal P}^D$ and ${\cal P}^{ND}$
are obtained from linear combinations of the diagrams of Fig.~3 corresponding
to various pseudoscalar mesons.

Combining Eq.s~\eno{qev}-\eno{piev} it is easy
to show that that the evolution equation for quark
distribution becomes
\bea
{d\over dt}(q+\bar q)&=&
(q+\bar q)\otimes \left(Q_{qq}+Q_{q\bar q}\right),
\label{tlnewp}\\
{d\over dt}(q-\bar q)&=&
(q-\bar q)\otimes \left(Q_{qq}- Q_{q\bar q}\right)+{\cal P}_{q\pi}\otimes 2
\pi,
\label{tlnewm}
\eea
where the various distributions and splitting functions are those which appear
in Eq.s~\eno{qev}-\eno{pidef}. It follows that the evolution equation of
Gottfried sum Eq.\eno{tlapsg} is formally
unchanged, provided of course the ${\cal Q}_{qq}$ and
${\cal Q}_{q\bar q}$ are the new splitting functions which include meson
emission.
Notice that the
meson distributions and splitting functions $\pi$, ${\cal P}_{\pi q}$
and ${\cal P}_{q\pi }$ do not directly contribute to the evolution of
$q+\bar q$, and thus of
$S_G$, whereas
they do contribute to the evolution of $q-\bar q$.\footnote{This
means that
the so-called Sullivan process, deep-inelastic scattering off quarks
inside pions in the nucleon wave function, cannot contribute directly to
measurements of $S_G$. This is in keeping with the naive observation that
a pion, being made of an equal number of $u$ and $d$ quarks (if quarks and
antiquarks are  counted with the same sign) cannot contribute to $S_G$.
Pion emission may, nevertheless, contribute {\it indirectly} to $S_G$ by
modifying the flavor structure of the sea. Thus whereas pion counting does not
tell us anything about $S_G$ (contrary to occasional naive statements) pion
emission mechanism (sometimes suggested within effective models\cite{pions})
may lead to a contribution to $S_G$.}
In order to avoid double counting the contributions to the QCD evolution
determined in this guise should not be added on top of the usual perturbative
evolution. Rather, at large enough $Q^2$ (presumably not larger than
$Q^2\sim 10$~GeV$^2$) the  nonperturbative evolution should flatten out,
so that at some large scale the perturbative behavior is reproduced.

The next required step is the computation of the splitting functions. Now, the
whole picture of $Q^2$ evolution as being described by a probabilistic process
relies on the dominance of certain diagrams (ladder diagrams) which contribute
to the deep-inelastic scattering process, and are then effectively resummed by
solving the evolution equations. This  corresponds physically
to the
dominance of processes where the emitted particle (a gluon in the
standard evolution equations, a meson in the present case)
is quasi-collinear to the original one (Weizs\"acker-Williams
approximation) and has
been proven\cite{DDT} to hold for gauge theories (both abelian and
nonabelian) in specific gauges in the Bjorken limit. These proofs can
actually be also
extended\cite{DDT} to the case of theories with ``pseudoscalar glue'',
i.e., to the case of coupling of fermions to point-like pseudoscalars. It is
much less clear that such a picture should hold in the case of extended
pseudoscalars: as a matter of fact, coupling to pseudoscalar glue for all
values of $Q^2$ would lead to scaling laws for the moments of structure
functions which are not in agreement with the experimental data \cite{altrev},
and the effects discussed here can only be important in a limited $Q^2$ range,
not so large that the quark-antiquark structure of the mesons is seen.

Hence, we proceed to the computation of the splitting function by assuming this
approximation to hold. We verify then that this assumption is consistent,
in that either significant $Q^2$ evolution occurs and
the meson emission cross
section is dominated by a kinematical range where the meson is
effectively pointlike, or, if the non-pointlike region is
providing an important contribution to the cross-section,
the evolution is negligible. This must happen at a value of $Q^2$ low enough
that the know asymptotic saling properties are unaffected.
The splitting functions are then given by
\beq
\left[{\cal P}_{q_iq_j}(x;t)\right]_\Pi=
\frac{d}{dt}\sigma^{\gamma^*\Pi}_{q_iq_j}(x;t),
\label{splitsig}
\eeq
in terms of the total cross section $\sigma^{\gamma^*\Pi}_{q_iq_j}(x;t)$
for the process of Fig.~3, integrated over all transverse momenta and expressed
in terms of the usual Bjorken variables. Using this definition in the evolution
equation \eno{tlapsg} the evolution of $S_G$ is immediately found to be given
by the multiplicative law
\beq
S_G(t)={\Delta}(t,t_0)S_G(t_0),
\label{qdel}
\eeq
where the nonperturbative determination of $\Delta(t,t_0)$
is
\bea
\left[{\Delta}_1(t,t_0)\right]_{\rm nonpert}&=
&\exp \bigg(
\left[{1\over6}\sigma^{\gamma^*\eta^\prime}_1(t)
+{1\over3}\sigma^{\gamma^*\eta^\prime}_1(t)
-{1\over 2}\sigma^{\gamma^*\pi}_1(t)\right]\nonumber\\
&  &\qquad\quad -\left[{1\over6}\sigma^{\gamma^*\eta^\prime}_1(t_0)
+{1\over3}\sigma^{\gamma^*\eta^\prime}_1(t_0)
-{1\over2}\sigma^{\gamma^*\pi}_1(t_0)\right]
\bigg).
\label{deldef}
\eea

The problem of determining the evolution of  $S_G$ is  thus reduced to the
computation of the various cross-sections which enter Eq.\eno{deldef}.
The computation, which is somewhat involved technically, has been performed and
discussed in Ref.\cite{bafogo}.
Because the pion-quark coupling is of course unknown,
the coupling has been written in the most general form compatible with Lorentz
invariance; this requires introducing four independent form factors (a
pseudoscalar, two axial, and a tensor). The cross-section is then computed from
the two diagrams (corresponding to $s$- and $t$-channel contributions) of
Fig.~4. Using the asymptotic behavior of the form factors for large values of
their arguments which is known from the behavior of the pion vertex function,
it is found that for intermediate values of $Q^2$ (up to a few GeV$^2$) the
cross sections are controlled essentially by the pseudoscalar form factor,
while the strength of one of the axial couplings controls the large $Q^2$
tail. The pseudoscalar form factor can be further constrained by use of the
Ward identity and a condition which relates it to the physical value of the
pion decay constant $f_\pi$. It turns out that all the evolution up
to intermediate $Q^2$  is thus
parametrized by a single parameter which can be identified with the
constituent quark mass, while an extra parameter (the axial coupling)
controls the large $Q^2$ tail.

The main qualitative features of the result of this computation can be
summarized as follows:
\begin{enumerate}
\item The cross sections are indeed dominated by the region where the
form factor is pointlike; this is the region where the emitted meson and quark
are almost collinear, consistently with the parton interpretation. This is
shown in Fig.5, where the cross section for pion emission is separated
into its various pieces (Fig.~5a), which are then displayed
(Fig.~5b,~5c) for various values of
$Q^2$ as a function of $x$. The pointlike dominance is apparent
by noting that the form factor in the $s$ and $t$ channel
are, respectively,
\hbox{$\varphi_{\hat s}=\varphi\big({1\over4} (2\hat s-M^2)\big)$}, and
\hbox{$\varphi_{\hat t}=\varphi\big({1\over4} (M^2-2\hat t)\big)$} with
\beq
\varphi^{\pi}(p^2)=\frac{m_d}{f_\pi}\frac{\Lambda^2+m_d^2}{\Lambda^2+p^2};
\label{ffac}
\eeq
and $s= {(1-x)\over x} Q^2$ as $x\to 1$, while $t=- {x\over 1-x} m^2_\pi$
in the collinear $k_{\rm perp}\to0$ limit, implying that the collinear,
pointlike limit is
$x\to1$ in the $s$ channel and $x\to 0$ in the $t$ channel.
It is interesting to observe that Fig.s~5b and 5c show that the growth of the
cross section with $Q^2$ (which determines, according to Eq.\eno{splitsig}, the
anomalous dimensions) is due to the growth of the collinear peak, which
dominates the cross section for all $Q^2$ larger than $\sim 0.1$~GeV$^2$.
In the very small $Q^2$ region ($Q^2 < 0.1$~GeV$^2$) where the non-pointlike
contributions are important the $Q^2$ dependence of the cross-section flattens
(Fig.~5a).

\item
The axial coupling controls only the high-$Q^2$ tail of  the
cross-section
(see Fig.~5a), while the bulk of the $Q^2$ dependence of the cross section
occurs in a region ($Q^2< 5$~GeV$^2$) where the cross section is controlled by
the pseudoscalar coupling.

\item
The anomalous dimensions are positive definite
throughout the $Q^2$ range.
The anomalous dimension for pion emission is indeed much larger than that
for eta emission, thus leading to screening of the Gottfried sum, according to
Eq.\eno{qdel},\eno{deldef}.
This is displayed in Fig.s~6a and 6b where the anomalous
dimension calculated from Eq.\eno{splitsig}
is shown in the extreme  cases of $\pi^0$ and $\eta^\prime$ emission.

\item
All anomalous dimensions flatten
at low $Q^2$ (Fig.~6), thus so does necessarily the evolution of $S_G$.
It follows that this can be smoothly connected  at the low scale
where flattening occurs to a quark
model predictions.

\item
All anomalous dimensions flatten
at large $Q^2$ (Fig.~6), due to the fact that
the cross section for meson emission behaves as an inverse power of $Q^2$ when
$Q^2$ is large enough. The evolution of $S_G$ flattens more rapidly than any of
the various evolutions due to meson emission separately does, since it is
proportional to a difference of mesonic cross sections, Eq.\eno{deldef}.
It follows that the mesonic effects disappear at large $Q^2$, where  they
behave as higher-twist effects,
and perturbative behavior is
regained.

\end{enumerate}

We proceed thus to a computation of the evolution of $S_G$
according to Eq.s~\eno{qdel}-\eno{deldef}.
The starting value of $S_G$ is chosen as the naive quark
model expectation Eq.\eno{naigot}. The scale at which this value holds is
fixed either as $|Q|=200$~MeV, or as the scale at which the evolution flattens.
Interestingly, this leads to the same result (small variations of the
flattening point being irrelevant due to the multiplicative nature of the
evolution).
The final result for the evolution of $S_G$ is shown
in Fig.7, for different values of the constituent quark mass (Fig.~7a)
and for different
values of the axial coupling which controls the high-$Q^2$ tail (fig.7b).
This shows that the experimental value is reproduced as a consequence of
nonperturbative evolution from the naive quark model value.
In the region where the data are taken or slightly above the
nonperturbative evolution flattens away; our computation provides an
effective smooth interpolation (which is however sensitive to a free
parameter of the computation, namely, the axial coupling)
between the nonperturbative evolution and
the perturbative behavior which holds in the high $Q^2$ limit.
 A robust prediction
of this computation is that significant evolution (characterized by anomalous
dimensions of several orders
of magnitude larger than the perturbative one) is
taking place in the intermediate $Q^2$ range, between roughly 0.1 and 10
GeV$^2$. In particular, for reasonable values of the axial coupling one
predicts that the asymptotic value of $S_G$  is yet
smaller than the presently reported one.

In conclusion, nonperturbative evolution of the Gottfried sum due to
bound-state
emission in an intermediate energy range $0.3\> \mbox{GeV}^2 \le |Q| \le 3$
GeV$^2$ is capable of reproducing the difference between the quark model
value and the experimental result taken at $Q^2 =4\>\mbox{GeV}^{2} $. The quark
model value holds at a low scale, and is reduced by nonperturbative generation
of a flavor-asymmetric sea, generated through a nonperturbative generalization
of the Altarelli-Parisi equations.
The required asymmetry follows from
the low-energy symmetry structure of QCD, as reflected in the spectrum
of pseudoscalar mesons, and is due to the anomalous breaking of flavor
singlet
chiral symmetry. In this sense, the data on
the Gottfried sum are displaying an effect of the axial anomaly.
Because the nonperturbative evolution flattens, the value of
$S_G$ thus obtained is preserved by further QCD evolution:
thus, nonperturbative QCD effects originating in the intermediate-energy region
leave a spur even in the asymptotic limit.

\section{The sigma term puzzle}

We are finally left with  assumption 3, Eq.\eno{sigpeq}.
The current knowledge of the
nucleon's structure functions is in
good agreement with this expectation: for example, at a scale of
Q$^2$=4~GeV$^2$ strange quarks carry about 2-3\%
of the nucleon's
momentum
\cite{strasea} (and accordingly much less at the nucleon's scale, since singlet
anomalous dimensions are generally large and in this case threshold effects
should also be relevant). This suggests that the matrix element
$\langle p|\bar s s | p\rangle$ should be roughly\footnote{
It can actually be shown rigorously\cite{forsig}
by defining parton distributions in
terms of QCD light-cone
operators that the forward matrix element of the operator
$\bar \psi_i \psi_i $ is to leading perturbative order equal to the first
moment of the distribution of quarks plus antiquarks of flavor $i$.}
of order 1 \% of
$\langle p|\bar u u | p\rangle$.
However, there is an old piece of data which seems to contradict this
conclusion; this is known
as the puzzle of the pion-nucleon sigma term\cite{sigpuz}.
The pion-nucleon sigma term is defined as the nucleon matrix element
of the light quark mass term in the QCD Hamiltonian:
\beq
\sigma \equiv \langle N | \bar{m} (\bar{u} u +\bar{d} d) | N \rangle
\label{sigdef}
\eeq
where $\bar m$ is the average light quark mass, i.e.,
$\bar{m} = (m_u +m_d )/2$.
The value of this operator can be extracted from experiment in two different
ways; it would seem that the only way the results of the two determination can
be made consistent with each other is by violating assumption 3. We show now,
however, that a more
careful analysis\cite{bafoti} reveals that this is not necessarily the
case.

The most direct way (at least theoretically) of determining the value of the
sigma term  is to observe that scalar quark densities appearing on the
r.h.s. of Eq.\eno{sigdef} can be obtained by commuting a pseudoscalar
charge with a pseudoscalar density according to
the basic
current algebra relation\cite{curral}
\beq
\left[\bar \psi\gamma^0\gamma_5 \tau^a\psi(x),\bar \psi\gamma_5 \tau^a\psi(y)
\right]=-\bar \psi\psi(x)\delta^{(4)}(x-y)
\label{currala}
\eeq
where  $\psi$ are quark isospinors and $\tau^a$ are Pauli isospin matrices.
The pseudoscalar current is in turn related to the pion interpolating
field $\phi_{\pi^a}$ by\cite{curral}
\beq
\partial_\mu \bar \psi\gamma^\mu\gamma_5 \tau^a\psi(x)=f_{\pi^a}
m^2_{\pi^a} \phi_{\pi^a},
\label{curralb}
\eeq
where $a$ labels the pion according to its isospin and $f_{\pi^a}$ and
$m_{\pi^a}$ are
the pion's decay constant and mass. Using Eq.s~\eno{currala}\eno{curralb}
the matrix element on the r.h.s. of Eq.\eno{sigdef} can be related through
straightforward current algebra manipulations\cite{curral}\cite{cheli} to a
pion-nucleon matrix element which, in turn, is related to the isospin-even
pion-nucleon scattering amplitude at vanishing momentum transfer.
This, up to several subtle
details of the extraction of the matrix element from the data, provides a
direct unambiguous determination of $\sigma$ as defined by Eq.\eno{sigdef},
with the result\cite{galesa}
\beq
\sigma = 45 \pm 10 \>\mbox{ MeV}.
\label{sigval}
\eeq

On the other hand, if we assume Eq.\eno{sigpeq}
(i.e. assumption 3) to hold,
we can equate the singlet and SU(3)-octet sigma terms:
\beq
\sigma \approx \sigma_{8} \equiv \bar{m} \langle N | (\bar{u} u + \bar{d} d
-2\bar{s} s ) | N \rangle.
\label{sigmoc}
\eeq
The reason why such an identification is interesting is that
$\sigma_{8} $ is proportional to the matrix element of the octet
portion ${\cal H}_{8} $ of the QCD Hamiltonian, which coincides with the octet
portion of the mass term since SU(3) flavor symmetry is only broken by masses
in QCD:
\bea
&&\sigma_{8} = \frac{3}{1- m_s /\bar{m} } \langle N | {\cal H}_{8} | N
\rangle
\label{sigh}\\
&&\quad {\cal H}_8 ={1\over3}\left(\bar m- m_s\right)
\left(\bar{u}u+\bar{d}d-2\bar{s}s\right).
\label{hdef}
\eea
Because in turn the matrix element of the octet part of the
Hamiltonian provides SU(3) breaking to first order in perturbation theory,
it can be measured from SU(3) mass splittings, for example in the baryon octet.

Indeed, it is well known that in the nonrelativistic quark model
one assumes
the mass splittings in the baryon octet to
be given by first order in perturbation by
an operator ${\cal O}_{8} $ which transforms as an octet under flavor SU(3)
(identified with the octet component of the strong
Hamiltonian). Group theory then leads to \cite{cheli}
\beq
M_B = M_0 + M_1 \langle B | Y | B \rangle + M_2 \langle B |
\left[I(I+1)-{1\over 4}Y^2 \right] | B \rangle,
\label{gmo}
\eeq
where $I$ and $Y$, respectively, are isospin and hypercharge
operators.
Eq.\eno{gmo} expresses the masses of the all the octet baryons in terms of the
three free
parameters $M_i $ and is in excellent agreement with experiment; in particular
the Gell-Mann--Okubo mass formula, which is a consequence of it, is one of the
great successes of the quark model.
The nucleon matrix element of the symmetry
breaking operator ${\cal O}_{8}$ can also be determined from
Eq.\eno{gmo} using SU(3) algebra:
\beq
\langle N | {\cal O}_{8} | N \rangle = M_1 - M_2 /2
=M_{\Lambda } - M_{\Xi } = (2M_N -M_{\Xi } -M_{\Sigma } )/3.
\label{sigmas}
\eeq
Identifying thus the symmetry breaking operator
${\cal O}_{8} $ with the octet component of the QCD
Hamiltonian ${\cal H}_{8} $, and using the known \cite{gale}
value of the (current) quark mass ratio $m_s /\bar{m}
=25 \pm 5 $, Eq.s~\eno{sigh},\eno{sigmas}
determine the octet sigma term as $\sigma_{8} = 25 \pm 5 $ MeV,
in blatant contradiction with the value
of $\sigma$ Eq.\eno{sigval}, i.e. with the identification
of $\sigma$ and $\sigma_8$ Eq.\eno{sigmoc}. This discrepancy is known as the
sigma term puzzle.

The most simple-minded conclusion, namely that $\sigma_8$ is much smaller than
$\sigma$ just because the nucleon matrix element
of $\bar s s $ is large leads to rather dramatic (and hard to believe)
consequences: because the strange quark is so much heavier than the light ones,
it would follow that about one third of the nucleon's mass is carried by
strange quarks, in contradiction to the parton model results \cite{strasea}
discussed above.
Alternative explanations, based on a failure of first-order perturbation theory
(in which case it's Eq.\eno{gmo} and its consequences which do not hold) are
extremely hard to reconcile with the success of the corresponding quark model
formulas. It is interesting to observe that, once again, effective models of
the nucleon have no difficulty in accommodating this state of affairs:
in the Skyrme model \cite{dona}
indeed assumption 3 is violated and $\sigma_8$ and $\sigma$
differ because of a large strange contribution, whereas in the bag
model\cite{jaf}
assumption 3 is  valid but first-order perturbation theory fails.

Upon closer
examination\cite{bafoti}, however, the argument given above is seen to contain
a
loophole: even though it is a true empirical fact that octet
mass splittings are
accurately described by the first order perturbative formula Eq.\eno{gmo}, how
do we know that the operator ${\cal O}_8$ which appears in this formula is the
same as the operator ${\cal H}_{8}$ which is related to the sigma term by
Eq.\eno{sigh}? Of course, classically this is necessarily the case: mass
splittings are given by the symmetry-breaking portion of the Hamiltonian, and
that's what ${\cal H}_{8}$ is. However in a second-quantized field theory this
need not necessarily be the case: quantum corrections may change the
symmetry structure, in general, hence also the pattern of
symmetry breaking (through a quantum anomaly). Now, it is well known that this
is precisely what happens for masses in the strong interactions: the masses
which enter quark model formulas are constituent quark masses, which
for light quarks differ by two orders of magnitude from the current quark
masses which appear in the fundamental Lagrangian (and Hamiltonian) of the
theory. Hence, what ${\cal O}_8$ should be identified with is the octet piece
of the {\it constituent} quark mass term, i.e., with the octet piece of
the effective low energy Hamiltonian of QCD\cite{georgi},
which is generally quite different from
the fundamental QCD Hamiltonian. Thus we have the possibility of escaping the
above dilemma if it just so happens that the octet portion of constituent
quark masses and current quark masses are different: in such a case
Eq.\eno{sigmoc}
still holds, along with first-order perturbation theory and its quark model
consequences, but, at least in principle, octet splittings calculated from
${\cal O}_8$  could be smaller than those calculated from
${\cal H}_8$, which would explain the above discrepancy without invoking a
violation of assumption 3.

This explanation is actually less contrived than it
may seem at first. We can see this by noting that actually the relationship
between the matrix elements of ${\cal O}_8$  and ${\cal H}_8$ is an exact
consequence of the so-called conformal anomaly equation. This is an expression
for the trace of the energy-momentum tensor $T^{\mu\nu}$
which is true at the operator level
in QCD:
\beq
T^\mu{}_\mu=(1+\gamma_m)
\sum_i m_i\bar\psi_i\psi_i+
{\beta(\alpha_s)\over4\alpha_s}
G^{\mu\nu}_aG_{\mu\nu}^a,
\label{confan}
\eeq
where the sum runs over all quark flavors,  $G^{\mu\nu}_a$ is the gluon field
strength,
$\beta={d\alpha_s(\mu )\over d\ln\mu }$ is the beta-function
for the strong coupling $\alpha_s={g^2\over4\pi}$, and
$\gamma_m(\mu )= -{d\ln m(\mu )\over d\ln\mu}$
is the mass anomalous dimension.
The last term  on the r.h.s. of Eq.\eno{confan} is
due to the conformal anomaly;
the anomalous dimension $\gamma_m$ is present because
the anomaly term and the mass term in Eq.\eno{confan} are not separately
scale invariant (i.e., they mix upon renormalization),
while the energy momentum tensor is
(up to surface terms). Classically, both $\gamma_m$ and the gluonic
contributions would be absent.

Taking the baryon matrix element of
Eq.\eno{confan} and using Lorentz invariance\cite{curral} (i.e., essentially
the fact that the trace of the energy-momentum tensor is a measure of
the violation of scale invariance) it follows  that the baryon mass
$M_B$ is given by
\beq
\langle B|\left(1+\gamma_m\right)\sum_i m_i\bar\psi_i\psi_i
 +{\beta(\alpha_s)\over4\alpha_s}
G^{\mu\nu}_aG_{\mu\nu}^a
 |B\rangle=M_B.
\label{wconf}
\eeq
This shows explicitly that indeed mass splittings are
linear in the expectation value of a certain operator,
as in lowest order perturbation theory: because of the
conformal anomaly, the linear ``approximation'' is actually exact;
quantum corrections appear as an additive contribution to the naive
first order perturbative formula.
 However, the operator
which ought to be used is that on the r.h.s. of Eq.\eno{wconf}. Otherwise
stated,
Eq.\eno{wconf} shows that
\beq
{\cal O}_{8}=
\left[\left(1+\gamma_m\right)\sum_i m_i\bar\psi_i\psi_i
 +{\beta(\alpha_s)\over4\alpha_s}
G^{\mu\nu}_aG_{\mu\nu}^a\right]_{(8)},
\label{odef}
\eeq
where the subscript (8) indicates the SU(3) octet component of the operator in
square brackets.

It follows that the baryon mass $M_B$ is in general given by the sum
of a current contribution $M_B^C$ and a dynamical contribution $M_B^D$:
\bea
&&M_B=M_B^C+M_B^D;
\label{rgsep}
\\
&&\quad M_B^C=\langle B|\sum_i m_i\bar\psi_i\psi_i |B\rangle,
\label{cmass}
\\
&& \quad M_B^D=\langle B|
{\beta(\alpha_s)\over4\alpha_s}
G^{\mu\nu}_aG_{\mu\nu}^a + \gamma _m\sum_i m_i\bar\psi_i\psi_i |B\rangle.
\label{dmass}
\eea
The octet portion of ${\cal H}_{8}$ provides the current mass splittings
of $M_B^C$,
while the
octet portion of ${\cal O}_{8}$ provides the full splittings of $M_B$ and
it reduces to ${\cal H}_{8}$
only in the limit in which both $\gamma_m$ and the gluonic term can be
neglected.
It should be stressed that this statement is an exact consequence of QCD.

Superficially, one may think that the gluonic operator in Eq.\eno{odef}
should be flavor singlet and thus should not contribute to mass splittings.
In this case, one would be left with $\gamma_m$, which is however not quite
negligible, since it may be computed in perturbation theory with the
result\cite{gale}
$\gamma_m[Q=1\mbox{ GeV}]=0.27$.\footnote{Notice that if this were the only
effect of the conformal anomaly, the discrepancy discussed above would be
even larger, since this extra term would make ${\cal O}_{8}$ larger
than ${\cal H}_{8}$, rather
than smaller as required to explain the puzzle.} This, however, is not possible
since, as already mentioned, the gluonic contribution is not separately
renormalization-group invariant, thus, by mixing with the fermionic component
it acquires a nonsinglet portion. Indeed, Eq.\eno{rgsep} is the unique
renormalization-group invariant decomposition of the baryon mass.

In fact, we can give two separate qualitative
reasons which suggest instead that indeed $M_B^D$ should be anticorrelated to
$M_B^C$. The first reason is based on the observation that a Ward identity
argument implies that the {\it one-meson reducible} contributions to $M_B^C$
and $M_B^D$ must cancel:
\beq
(M_B^C)^{ omr}+(M_B^D)^{ omr}=0.
\label{wacanc}
\eeq
Here $M_B^{ omr}$ are defined as the contribution to the
matrix elements Eq.\eno{rgsep}-\eno{dmass}
where the
operators which enter the definitions Eq.\eno{cmass}\eno{dmass} couple to the
baryon state by first coupling to a meson which then couples to the baryon.
Because such contributions are expected (by pole-dominance arguments) to
provide the bulk of $M_B^D$, Eq.\eno{wacanc} suggests that $M_B^D$ has a
large component anticorrelated to $M_B^C$.

An independent reason is given by the observation that
Eq.\eno{wconf} can be viewed as  a relation between current masses
$m^C_q$ (given by Eq.\eno{cmass} assuming that the matrix element of
$\bar\psi_i\psi_i$ is just the number of quarks plus antiquarks of
flavor $i$)
and constituent quark
masses $m_q$
(which add up to
the baryon mass $M_B$); accordingly,
the decomposition Eq.\eno{rgsep}
determines the dynamical contribution to the constituent quark mass
(i.e., the difference between constituent and current mass)
as a function of the
current mass $m^C_q$:
\beq
m^D_q\equiv m_q(m)-m^C_q=m^D_q(m).
\label{dmdef}
\eeq
A baryon mass splitting will coincide with the mass splitting computed at the
current level, i.e.,  neglecting the dynamical contribution $M_B^D$
Eq.\eno{dmass}, only if $m^D_q$ Eq.\eno{dmdef} turns out to be a flat function,
i.e., not to depend on the current mass, so that the current and constituent
masses differ by a fixed, flavor independent amount.
However, the qualitative behavior of the
function $m_q^D(m)$ Eq.~\eno{dmdef}
is actually known. Indeed, when the current mass
vanishes, the constituent mass reduces to roughly 300~MeV, i.e., a third of the
mass of the nucleon, which is made of quasi-massless quarks; hence
$m_q^D(m=0)=300$~MeV. When the current
mass tends to infinity, quarks become quasi-static,
and the current and constituent masses coincide; hence
$\lim_{m\to\infty}m_q^D(m)=0$. It follows that
$m^D_q$
is a {\it decreasing}
function of the constituent mass, and not a flat function.
This implies that constituent mass splittings
are in general smaller than current mass splittings
(so, for instance, the strange-light splitting is {\it smaller}
for constituent masses than it is for current ones): the contribution of
$M_B^D$ to mass splittings is anticorrelated to that of $M_B^D$.

Both these qualitative arguments can actually be turned into
quantitative (or at least semi-quantitative) ones\cite{bafoti}.
Firstly, one can
actually estimate $(M_B^D)^{\rm omr}$ by pole-dominance arguments,
thus obtaining a rough estimate of the value of $M_B^D$. Also,
it is possible to actually compute the function
$m_q^D(m)$ by studying the full quark
propagator: more specifically, it is possible
to attempt  an approximate resolution of
the Schwinger-Dyson equation satisfied by the full quark propagator.
This leads to a determination of the quark self-energy, thus also of the quark
constituent mass; this in turn can be assumed to give the baryon mass by an
additive relation (i.e., the baryon mass is given by the sum of the masses of
its constituent quarks).
In either case, it is possible to determine $\Delta M$, defined as
\beq
\Delta M\equiv\langle N|{\cal O}_8|N\rangle  -\langle N|{\cal H}_8|
N\rangle
\label{deltamdef}.
\eeq
Even though the former determination is rather crude,
and the latter is affected by an underlying theoretical uncertainty due to
lack of knowledge of the QCD dynamics in the infrared (as reflected, for
example, by the behavior of the strong coupling in the infrared limit),
both  determinations lead to a value
$\Delta M \sim 150$~MeV
(with large uncertainties, perhaps up to 50 \%). Specifically,
with\cite{bafoti}
$30\>\mbox{MeV} \le \Delta M \le 250 \>\mbox{MeV}$, and using
Eq.\eno{deltamdef}
to relate
Eq.\eno{gmo} (mass splittings) to Eq.\eno{sigh}
(matrix elements of the sigma term)
leads to the value $30\>\mbox{MeV}\le \sigma_8\le 60\>\mbox{MeV}$, in
perfect agreement with the value of $\sigma$ Eq.\eno{sigval}, i.e.
with the identification of $\sigma$ and $\sigma_8$, consistently with
assumption 3 [Eq.\eno{sigpeq}].

In conclusion, if one takes into account that even ``gluonic'' quantities
such as the dynamical contribution to baryon masses can have a flavor
nonsinglet component, it follows that mass splittings at the constituent and
current level are not the same. This allows one to resolve the puzzle of the
discrepancy between the two different determinations of the sigma term, without
invoking a large strange content of the nucleon, and without violating
the linear quark model mass formulas, such as those leading to the
Gell-Mann--Okubo relation.
\section{Conclusions}

There are two main lessons to be learnt from this work, a phenomenological one
and a theoretical one.
The phenomenological lesson is that instances of discrepancy between
experimental data and expectations based on the quark model are resolved once
nonperturbative effects are taken into account.
The theoretical lesson is that these effects appear both in the QCD evolution
of operators, and in the computation of the operators themselves (in
particular, in the determination of their symmetry structure). Such
contributions, even though of course they originate in the gluon dynamics,
depend strongly on the quantum numbers of the physical states under
investigation and cannot be simply parametrized by universal flavor-singlet
parameters, such as vacuum condensates.
These two points, taken together, suggest that new data may actually lead us to
a better understanding of the infrared dynamics of QCD and thus to go  beyond
the very successful  but limited set of predictions which perturbative QCD
allows to make.

\bigskip
\noindent{\bf Acknowledgements:}
I thank the organizers of the school and especially M~A.~Nowak
for managing to assemble an unusually
stimulating set of talks and discussions.
This paper is largely based on work done in collaboration with
R.~D.~Ball\cite{bafogo}\cite{bafoti};
I also thank V.~Barone for discussions.
This paper is partially based on notes taken by
M.~Engelhardt and K.~Lucke (University of Erlangen) at a lecture given by me
at the Erlangen-Regensburg Graduiertenkolleg.
\vfill
\eject

\vfill\eject
\leftline{\Large \bf Figure Captions}
\smallskip
\medskip
\noindent{\bf Fig.~[1]}
The gluon radiation diagram which leads to one-loop
Altarelli-Parisi evolution.

\medskip
\noindent{\bf Fig.~[2]}
The two-loop diagram which generates
nonsinglet evolution
of quark distribution. When $i=j$ the final state must be antisymmetrized with
respect to the two identical quarks.

\medskip
\noindent{\bf Fig.~[3]}
The meson radiation diagram which generates nonperturbative
evolution.

\medskip
\noindent{\bf Fig.~[4]}
Deep-inelastic scattering off a quark which radiates a bound state
$\Pi$: a) $t$-channel diagram; b) $s$-channel diagram.

\medskip
\noindent{\bf Fig.~[5]}
The $\pi^0$ emission cross section (with constituent quark mass
$M=325$ MeV) computed from the two diagrams of Fig.4: a)
cross section integrated over $x$ as a function of $Q^2$; full line: full cross
section with no axial coupling; dot-dash line:
full cross
section with axial coupling; dashed  line: $t$-channel contribution (no axial
coupling), diagram Fig.4a;  dotted line: $s$-channel contribution (no axial
coupling), diagram Fig.4b. b) $t$-channel contribution (dashed line of Fig.5a)
as a function of $x$ and $Q^2$. c) $s$-channel contribution (dotted
line of Fig.5a)
as a function of $x$ and $Q^2$.

\medskip
\noindent{\bf Fig.~[6]}
Anomalous dimension calculated from Eq.\eno{splitsig}
for meson emission, for different values of the constituent quark mass:
a)~$\pi^0$ emission; b)~$\eta^\prime$ emission. The axial coupling is set
to a large (i.e. maximal) value.

\medskip
\noindent{\bf Fig.~[7]}
Scale dependence of $S_G$ computed from Eq.\eno{qdel}:
a) variation with the constituent quark mass; the axial coupling
is set to a maximal value; b) $M_q= 350$ MeV; full curve: maximal axial
coupling; dashed curve: no axial coupling.
The experimental point Eq.\eno{gotexpa}
is also shown.
\vfill

\begin{thebibliography}{99}
\bibitem{EMC} J.~Ashman et al., Phys. Lett. {\bf B206 }, 364  (1988);
Nucl. Phys. {\bf B328}, 1 (1990).
\bibitem{spinrev} For a review see G.~Altarelli, in ``The Challenging
Questions'', Proc. of the 1989 Erice School, A.~Zichichi, ed. (Plenum, New
York, 1990).
\bibitem{forspin}  Phys. Lett.
{\bf B224 }, 189  (1989); S.~Forte, Nucl. Phys. {\bf B 331}, 1 (1990);
S.~Forte and E.~V.~Shuryak, Nucl. Phys. {\bf B 357}, 153 (1991).
\bibitem{NMCa}  P.~Amaudruz et al., Phys. Rev. Lett. {\bf 66},
2712 (1991).
\bibitem{NMCb} P.~Amaudruz et al., CERN preprint CERN-PPE/93-117 (1993).
\bibitem{SSV}  T.~Sloan, G.~Smadja and R.~Voss, Phys. Rep. {\bf 162 }, 45
(1988).
\bibitem{sigpuz}  For a review see e.g. R.~L.~Jaffe and C.~L.~Korpa, Comm.
Nucl. Part. Phys., {\bf 17} 163 (1987).
\bibitem{altrev} G.~Altarelli, Phys. Rep. {\bf 81}, 1 (1982).
\bibitem{soper} See e.g. J.~C.~Collins, D.~E.~Soper and G.~Sterman, in
``Perturbative Chromodynamics'', A.~H.~Mueller ed. (World Scientific,
Singapore, 1989).
\bibitem{NMCrat}  P.~Amaudruz et al., Nucl. Phys. {\bf B371}, 3  (1992).
\bibitem{leapre}  see
e.g. E.~Leader and E.~Predazzi, ``Gauge Theories and the `New Physics' ''
(Cambridge University, Cambridge, 1982).
\bibitem{GSR}  K.~Gottfried, Phys. Rev. Lett. {\bf 18}, 1174 (1967)
\bibitem{shad} V.~R.~Zoller, Phys. Lett.  {\bf B279}, 145  (1992); B.~Bade\l ek
and J.~Kwieci\'nski, Nucl. Phys. {\bf B370}, 178 (1991).
\bibitem{gotfor} S.~Forte, Phys. Rev.  {\bf D47 }, 1842  (1993).
\bibitem{forsig} M.~Anselmino and S.~Forte, Torino preprint DFTT 6/92 (1993),
Z.~Phys.~C, in press.
\bibitem{rosac} D.~A.~ Ross, C.~T.~Sachrajda, Nucl. Phys. {\bf B 149}, 497
(1979).
\bibitem{ehq} E.~J.~Eichten, I.~Hinchliffe and
C.~Quigg,  Phys. Rev. {\bf D45}, 2269 (1992).
\bibitem{bafogo} R.~D.~Ball, S.Forte, Oxford and Torino preprint
OUTP-93-18P and DFTT 9/93 (1993).
\bibitem{pions} A.~W.~Thomas, Nucl. Phys. {\bf A532}, 177 (1991);
S.~Kumano, Phys. Rev. {\bf D43}, 59 (1991); {\bf D43}, 3067 (1991);
S.~Kumano and J.~T.~Londergan, Phys. Rev. {\bf D44},717 (1991).
\bibitem{DDT} See e.g. Y.~L.~Dokshitzer, D.~I.~Dyakonov and S.~I.~Troyan,
Phys. Rep. {\bf 58}, 269 (1980)
\bibitem{strasea} See V.~Barone et al., Zeit. Phys. {\bf C58}, 541 (1993);
A.~D.~Martin, W.~J.~Stirling and R.~G.~Roberts, Phys. Rev. {\bf D47}, 867
(1993).
\bibitem{bafoti} R.~D.~Ball, S.~Forte and J.Tigg, Oxford and Torino preprint
OUTP-92-35P and DFTT 69/92 (1993)
\bibitem{curral} See e.g. V.~de Alfaro, S.~Fubini, G.~Furlan and C.~Rossetti,
``Currents in Hadron Physics'' (North-Holland, Amsterdam, 1973).
\bibitem{cheli} T.-P.~Cheng and L.-F.~Li, ``Gauge Theories of Elementary
Particle Physics'' (Clarendon, Oxford, 1984).
\bibitem{galesa} J.~Gasser, H.~Leutwyler
and M.~E.~Sainio,  Phys. Lett. {\bf B253}, {252} (1991).
\bibitem{gale} See e.g. J.~Gasser and H.~Leutwyler,
Phys. Rep., {\bf 87} 77 (1982).
\bibitem{dona} J.~F.~Donoghue and C.~R.~Nappi,
Phys. Lett. {\bf 168B}, {105} (1986).
\bibitem{jaf} R.~L.~Jaffe, Phys. Rev. {\bf D21}, 3215 (1980).
\bibitem{georgi} See e.g. H.~Georgi ``Weak Interactions and Modern Particle
Theory'' (Benjamin-Cummings, Menlo Park, CA, 1984).
\end{thebibliography}
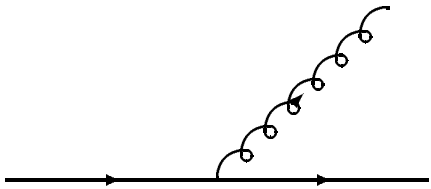
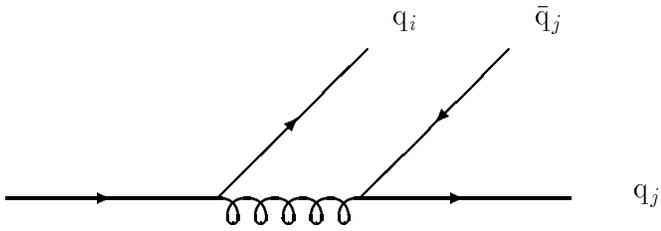
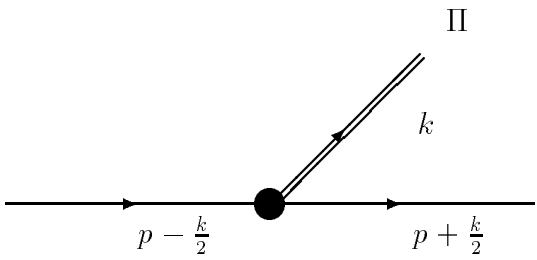
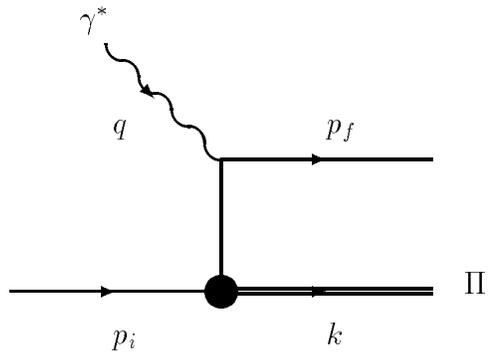
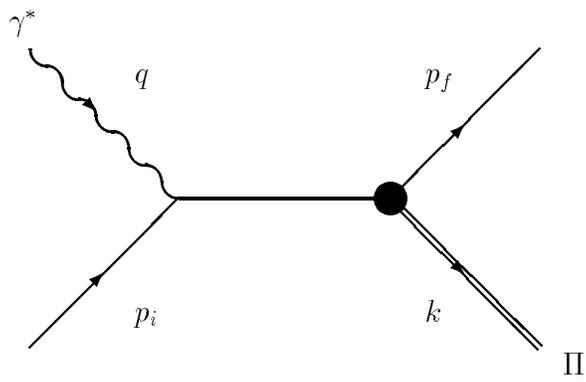
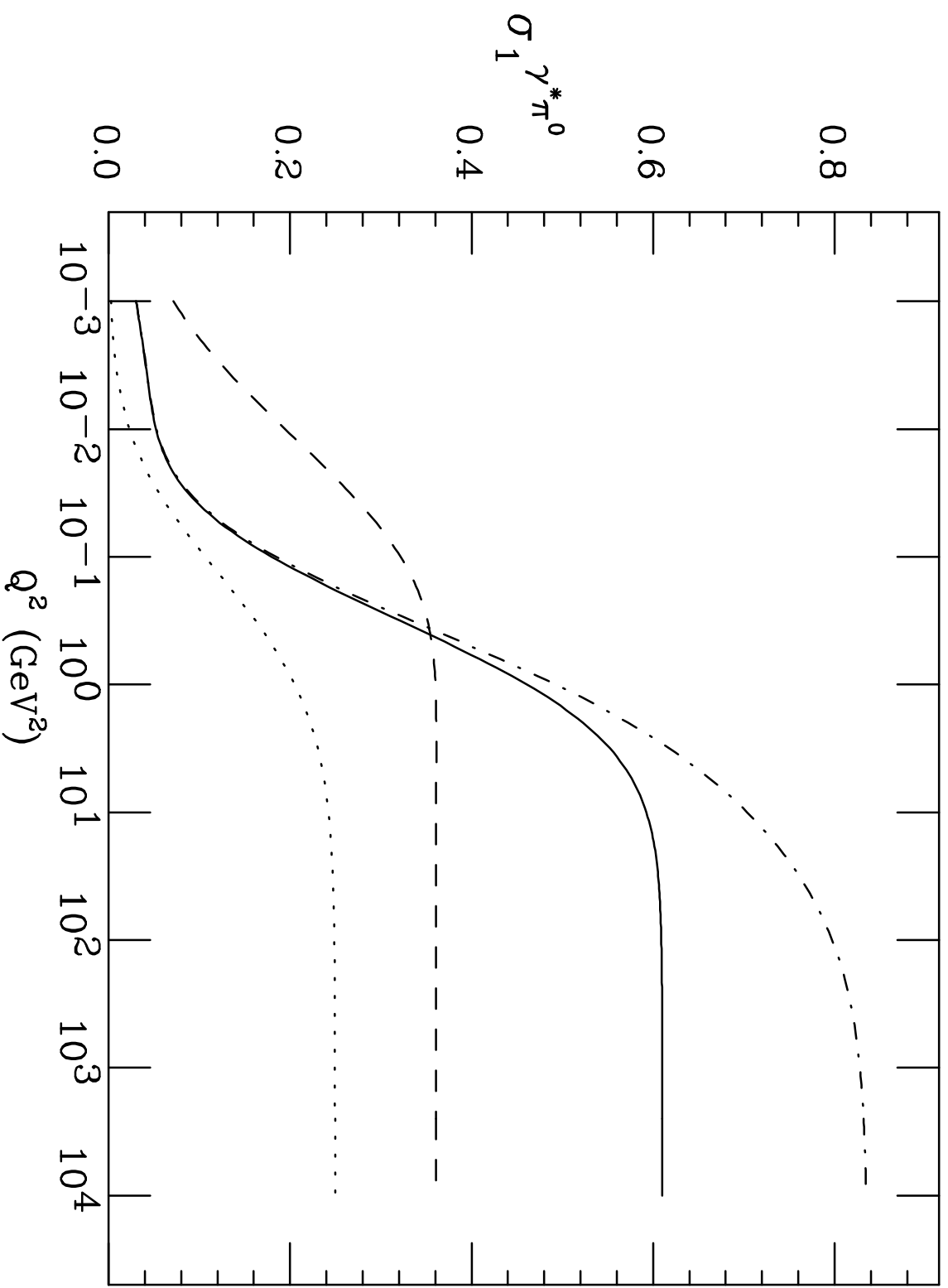
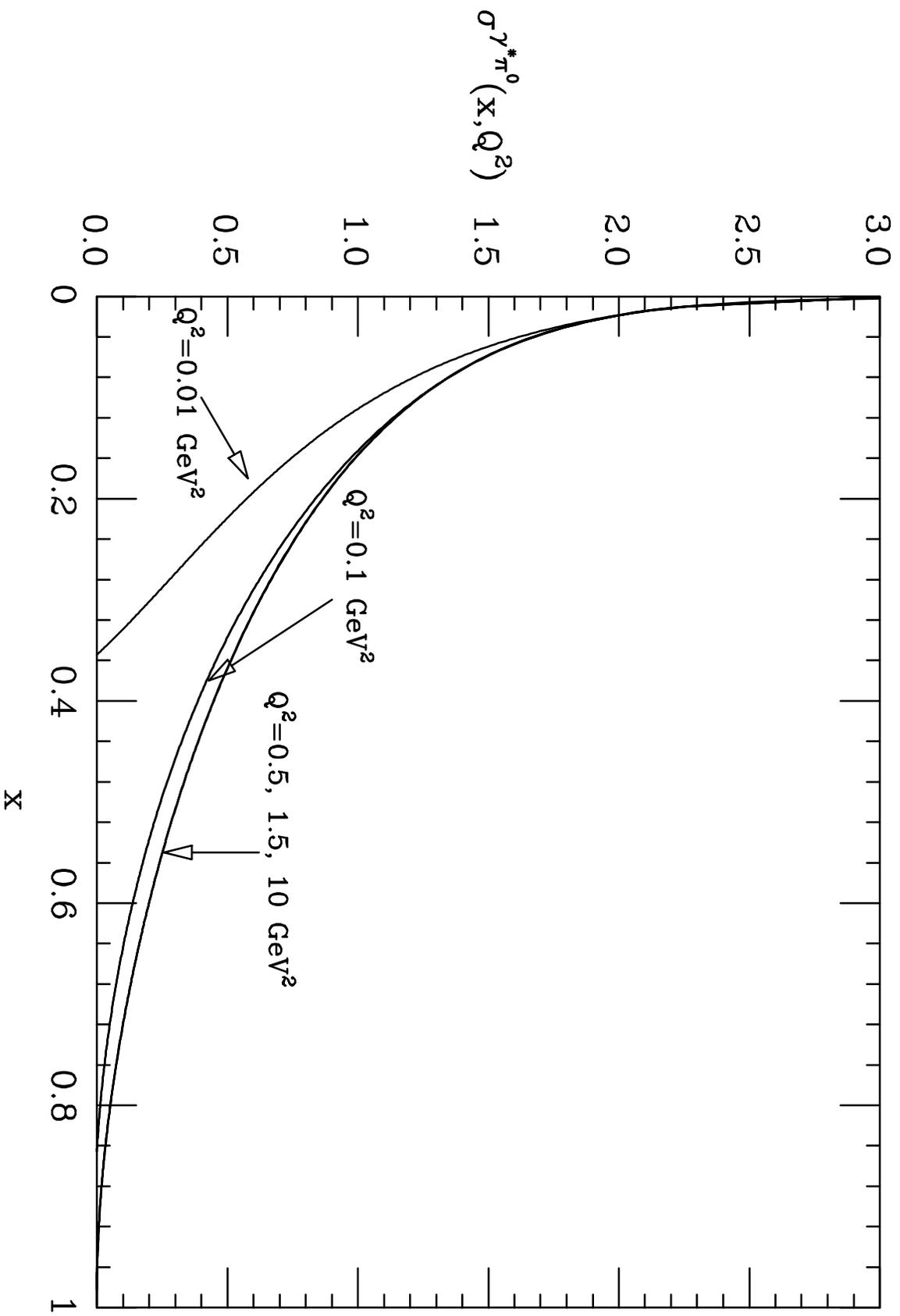
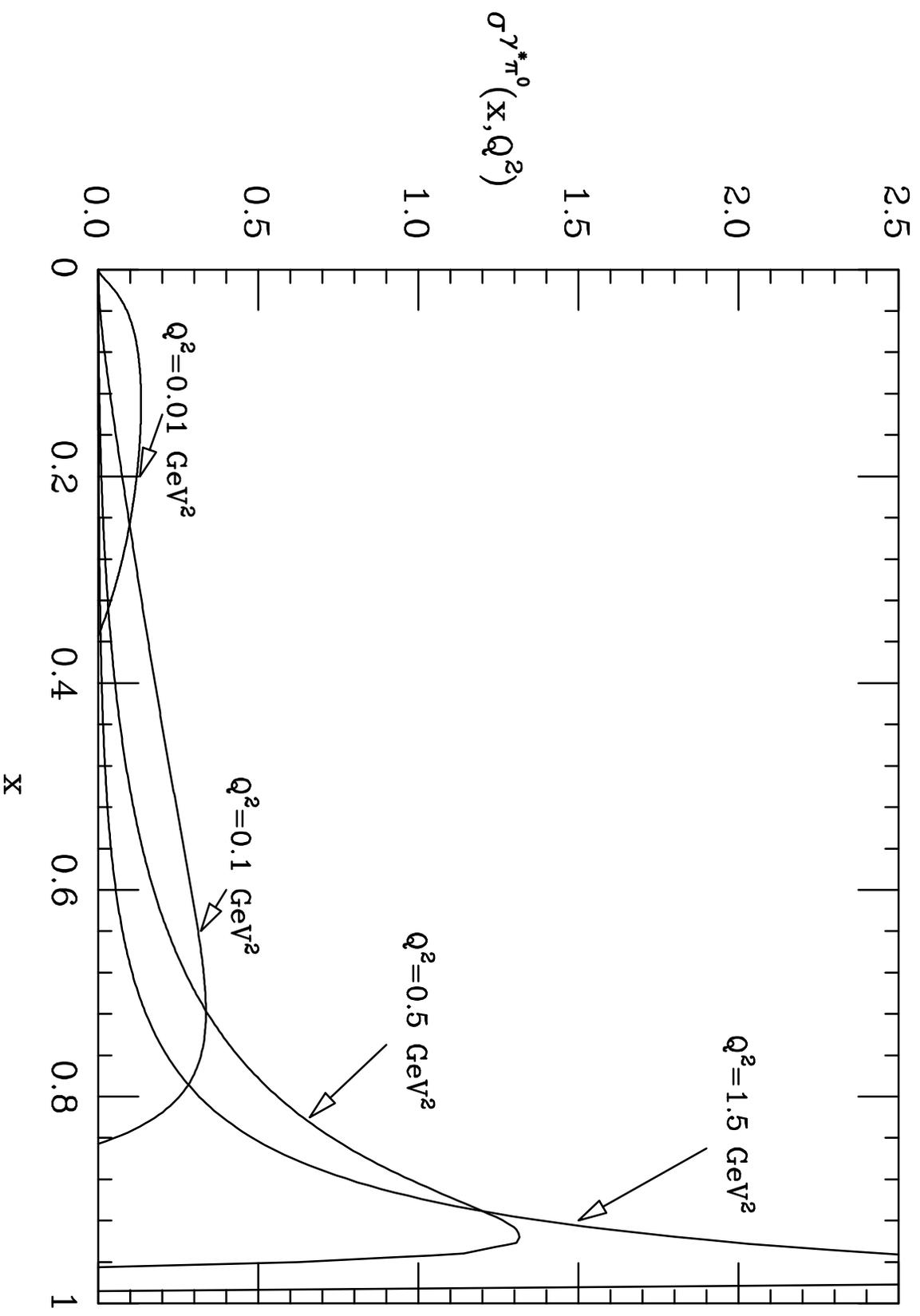
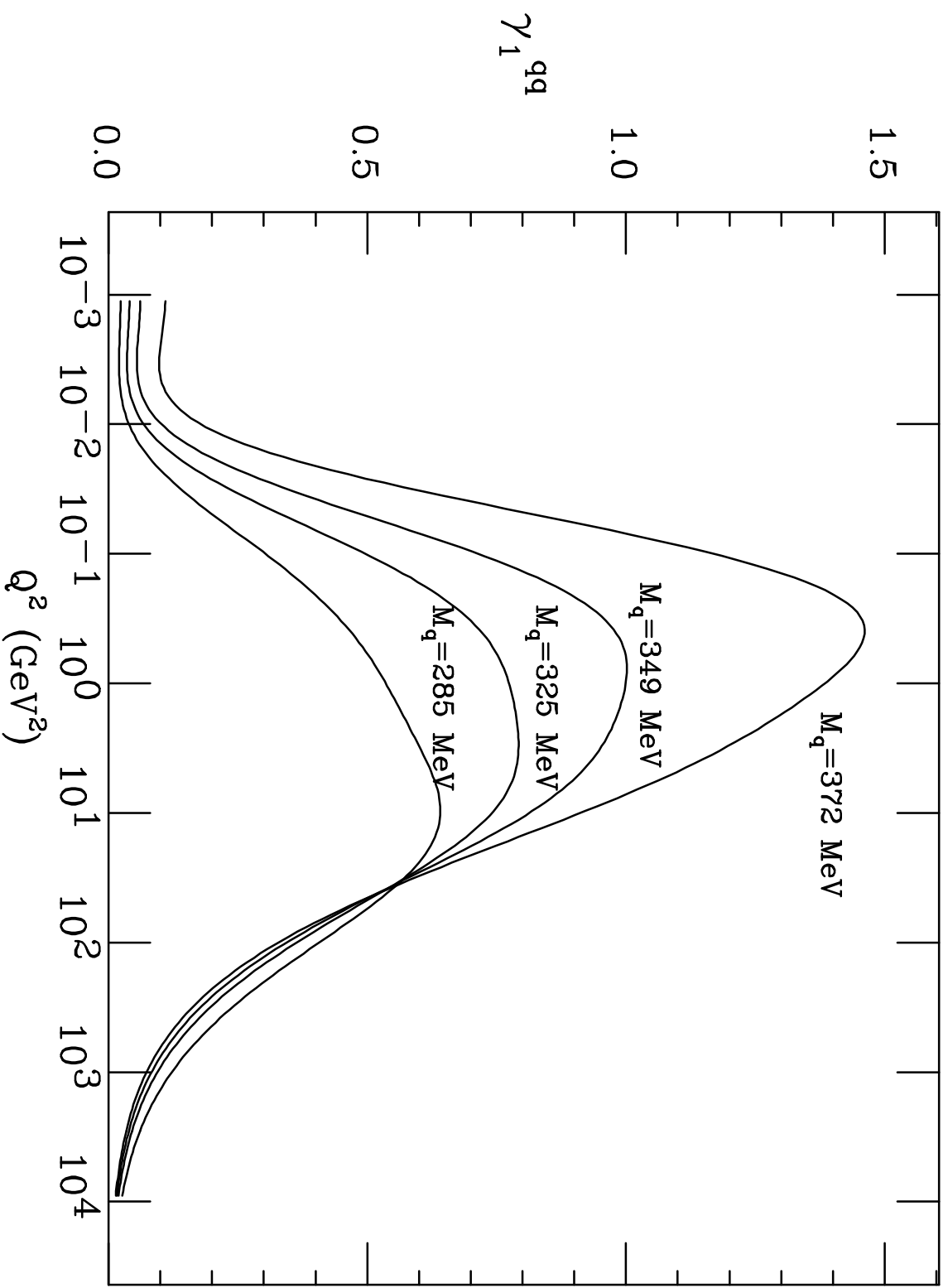
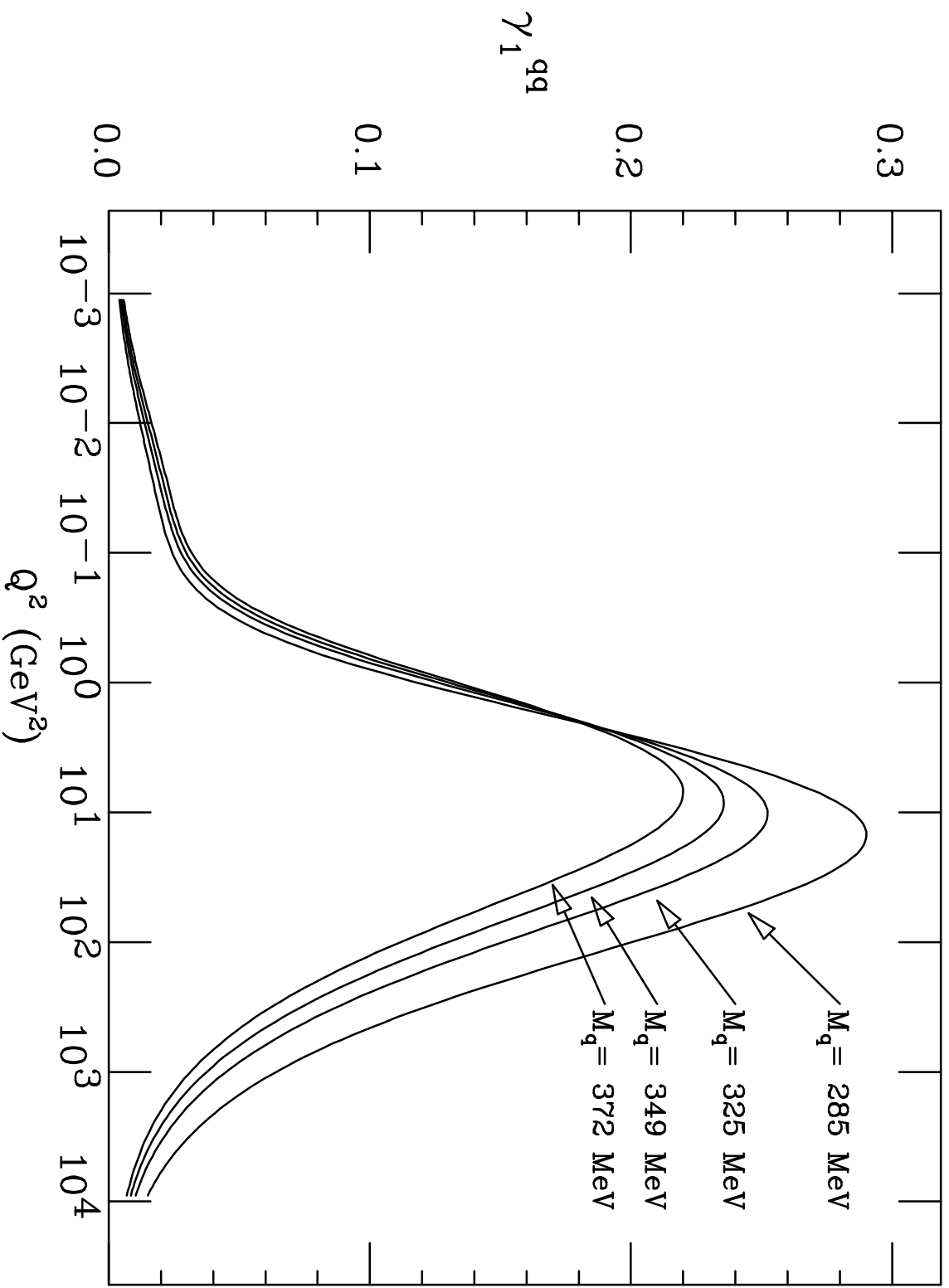
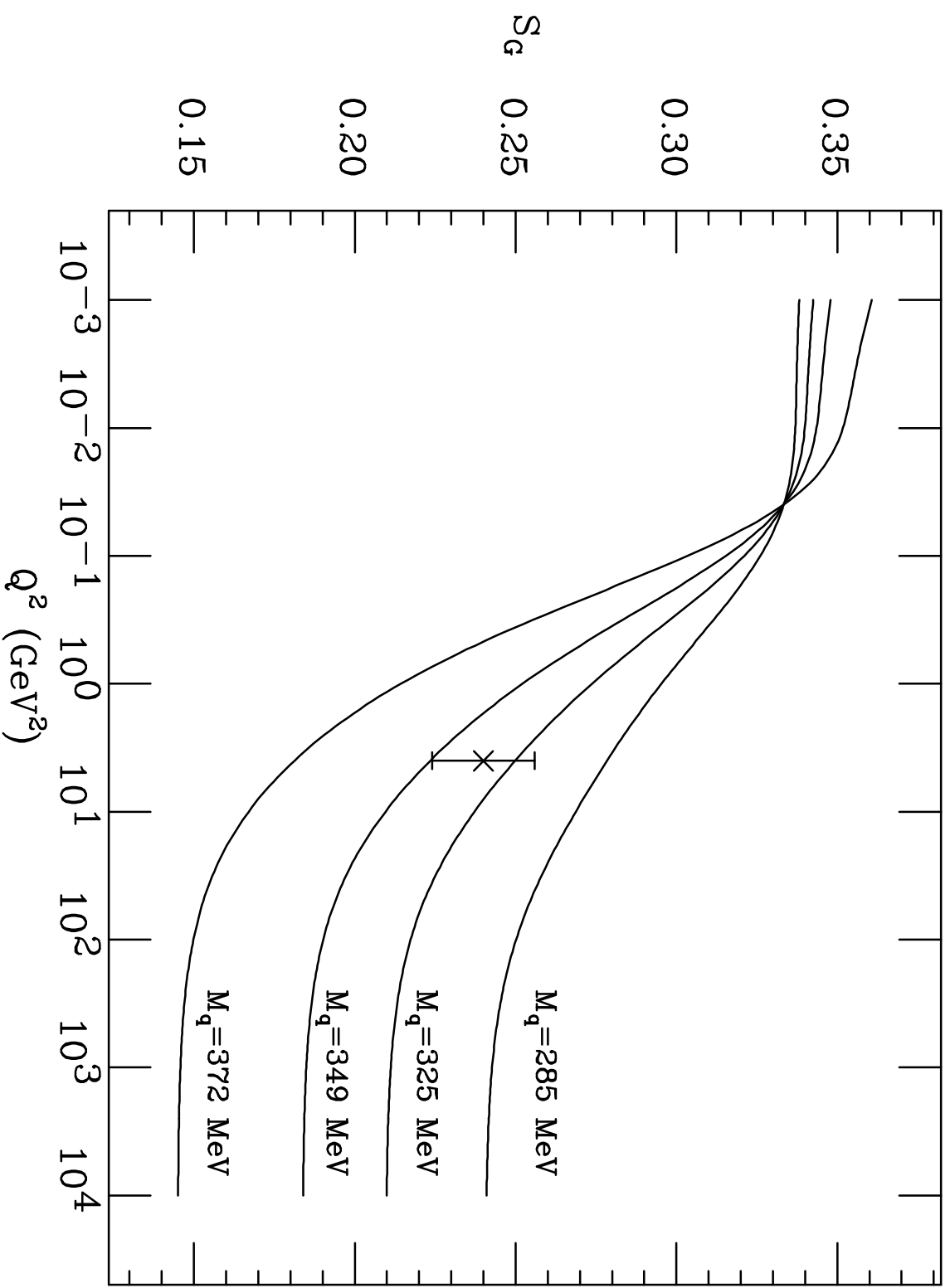
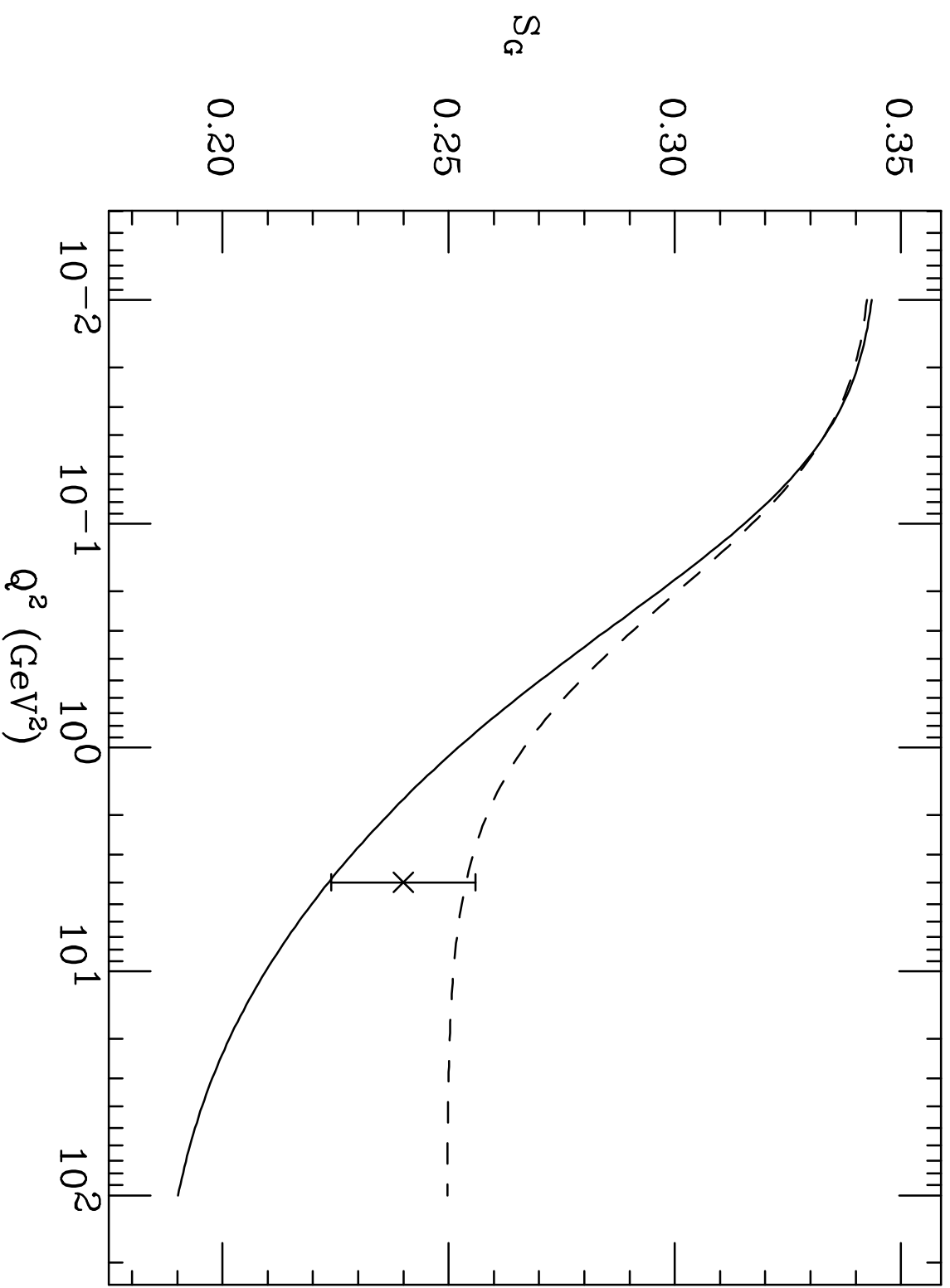
\end{document}